\overfullrule=0pt
\input harvmac


\edef\resetatcatcode{\catcode`\noexpand\@\the\catcode`\@\relax}

\def\makeatletter{\catcode`\@11\relax}

\makeatletter

\ifx\miniltx\undefined\else \fi
\let\miniltx\box

\def\@makeother#1{\catcode`#1=12\relax}

\def\@ifnextchar#1#2#3{%
  \let\reserved@d=#1%
  \def\reserved@a{#2}\def\reserved@b{#3}%
  \futurelet\@let@token\@ifnch}
\def\@ifnch{%
  \ifx\@let@token\@sptoken
    \let\reserved@c\@xifnch
  \else
    \ifx\@let@token\reserved@d
      \let\reserved@c\reserved@a
    \else
      \let\reserved@c\reserved@b
    \fi
  \fi
  \reserved@c}
\begingroup
\def\:{\global\let\@sptoken= } \:  
\def\:{\@xifnch} \expandafter\gdef\: {\futurelet\@let@token\@ifnch}
\endgroup

\def\@ifstar#1{\@ifnextchar *{\@firstoftwo{#1}}}
\long\def\@dblarg#1{\@ifnextchar[{#1}{\@xdblarg{#1}}}
\long\def\@xdblarg#1#2{#1[{#2}]{#2}}

\long\def \@gobble #1{}
\long\def \@gobbletwo #1#2{}
\long\def \@gobblefour #1#2#3#4{}
\long\def\@firstofone#1{#1}
\long\def\@firstoftwo#1#2{#1}
\long\def\@secondoftwo#1#2{#2}

\def\NeedsTeXFormat#1{\@ifnextchar[\@needsf@rmat\relax}
\def\@needsf@rmat[#1]{}
\def\ProvidesPackage#1{\@ifnextchar[%
    {\@pr@videpackage{#1}}{\@pr@videpackage#1[]}}
\def\@pr@videpackage#1[#2]{\wlog{#1: #2}}

\let\DeclareOption\@gobbletwo
\def\ProcessOptions{\@ifstar\relax\relax}

\def\RequirePackage{%
  \@fileswithoptions\@pkgextension}
\def\@fileswithoptions#1{%
  \@ifnextchar[
    {\@fileswith@ptions#1}%
    {\@fileswith@ptions#1[]}}
\def\@fileswith@ptions#1[#2]#3{%
  \@ifnextchar[
  {\@fileswith@pti@ns#1[#2]#3}%
  {\@fileswith@pti@ns#1[#2]#3[]}}

\def\@fileswith@pti@ns#1[#2]#3[#4]{%
    \def\reserved@b##1,{%
      \ifx\@nil##1\relax\else
        \ifx\relax##1\relax\else
         \noexpand\@onefilewithoptions##1[#2][#4]\noexpand\@pkgextension
        \fi
        \expandafter\reserved@b
      \fi}%
      \edef\reserved@a{\zap@space#3 \@empty}%
      \edef\reserved@a{\expandafter\reserved@b\reserved@a,\@nil,}%
  \reserved@a}

\def\zap@space#1 #2{%
  #1%
  \ifx#2\@empty\else\expandafter\zap@space\fi
  #2}

\let\@empty\empty
\def\@pkgextension{sty}

\def\@onefilewithoptions#1[#2][#3]#4{%
  \input #1.#4 }

\def\typein{%
  \let\@typein\relax
  \@testopt\@xtypein\@typein}
\def\@xtypein[#1]#2{%
  \message{#2}%
  \advance\endlinechar\@M
  \read\@inputcheck to#1%
  \advance\endlinechar-\@M
  \@typein}
\def\@namedef#1{\expandafter\def\csname #1\endcsname}
\def\@nameuse#1{\csname #1\endcsname}
\def\@cons#1#2{\begingroup\let\@elt\relax\xdef#1{#1\@elt #2}\endgroup}
\def\@car#1#2\@nil{#1}
\def\@cdr#1#2\@nil{#2}
\def\@carcube#1#2#3#4\@nil{#1#2#3}
\def\@preamblecmds{}

\def\@star@or@long#1{%
  \@ifstar
   {\let\l@ngrel@x\relax#1}%
   {\let\l@ngrel@x\long#1}}

\let\l@ngrel@x\relax
\def\newcommand{\@star@or@long\new@command}
\def\new@command#1{%
  \@testopt{\@newcommand#1}0}
\def\@newcommand#1[#2]{%
  \@ifnextchar [{\@xargdef#1[#2]}%
                {\@argdef#1[#2]}}
\long\def\@argdef#1[#2]#3{%
   \@ifdefinable #1{\@yargdef#1\@ne{#2}{#3}}}
\long\def\@xargdef#1[#2][#3]#4{%
  \@ifdefinable#1{%
     \expandafter\def\expandafter#1\expandafter{%
          \expandafter
          \@protected@testopt
          \expandafter
          #1%
          \csname\string#1\expandafter\endcsname
          {#3}}%
       \expandafter\@yargdef
          \csname\string#1\endcsname
           \tw@
           {#2}%
           {#4}}}
\def\@testopt#1#2{%
  \@ifnextchar[{#1}{#1[#2]}}
\def\@protected@testopt#1{
  \ifx\protect\@typeset@protect
    \expandafter\@testopt
  \else
    \@x@protect#1%
  \fi}
\long\def\@yargdef#1#2#3{%
  \@tempcnta#3\relax
  \advance \@tempcnta \@ne
  \let\@hash@\relax
  \edef\reserved@a{\ifx#2\tw@ [\@hash@1]\fi}%
  \@tempcntb #2%
  \@whilenum\@tempcntb <\@tempcnta
     \do{%
         \edef\reserved@a{\reserved@a\@hash@\the\@tempcntb}%
         \advance\@tempcntb \@ne}%
  \let\@hash@##%
  \l@ngrel@x\expandafter\def\expandafter#1\reserved@a}
\long\def\@reargdef#1[#2]#3{%
  \@yargdef#1\@ne{#2}{#3}}
\def\renewcommand{\@star@or@long\renew@command}
\def\renew@command#1{%
  {\escapechar\m@ne\xdef\@gtempa{{\string#1}}}%
  \expandafter\@ifundefined\@gtempa
     {\@latex@error{\string#1 undefined}\@ehc}%
     {}%
  \let\@ifdefinable\@rc@ifdefinable
  \new@command#1}
\long\def\@ifdefinable #1#2{%
      \edef\reserved@a{\expandafter\@gobble\string #1}%
     \@ifundefined\reserved@a
         {\edef\reserved@b{\expandafter\@carcube \reserved@a xxx\@nil}%
          \ifx \reserved@b\@qend \@notdefinable\else
            \ifx \reserved@a\@qrelax \@notdefinable\else
              #2%
            \fi
          \fi}%
         \@notdefinable}
\let\@@ifdefinable\@ifdefinable
\long\def\@rc@ifdefinable#1#2{%
  \let\@ifdefinable\@@ifdefinable
  #2}
\def\newenvironment{\@star@or@long\new@environment}
\def\new@environment#1{%
  \@testopt{\@newenva#1}0}
\def\@newenva#1[#2]{%
   \@ifnextchar [{\@newenvb#1[#2]}{\@newenv{#1}{[#2]}}}
\def\@newenvb#1[#2][#3]{\@newenv{#1}{[#2][#3]}}
\def\renewenvironment{\@star@or@long\renew@environment}
\def\renew@environment#1{%
  \@ifundefined{#1}%
     {\@latex@error{Environment #1 undefined}\@ehc
     }{}%
  \expandafter\let\csname#1\endcsname\relax
  \expandafter\let\csname end#1\endcsname\relax
  \new@environment{#1}}
\long\def\@newenv#1#2#3#4{%
  \@ifundefined{#1}%
    {\expandafter\let\csname#1\expandafter\endcsname
                         \csname end#1\endcsname}%
    \relax
  \expandafter\new@command
     \csname #1\endcsname#2{#3}%
     \l@ngrel@x\expandafter\def\csname end#1\endcsname{#4}}

\def\providecommand{\@star@or@long\provide@command}
\def\provide@command#1{%
  {\escapechar\m@ne\xdef\@gtempa{{\string#1}}}%
  \expandafter\@ifundefined\@gtempa
    {\def\reserved@a{\new@command#1}}%
    {\def\reserved@a{\renew@command\reserved@a}}%
   \reserved@a}%

\def\@ifundefined#1{%
  \expandafter\ifx\csname#1\endcsname\relax
    \expandafter\@firstoftwo
  \else
    \expandafter\@secondoftwo
  \fi}

\chardef\@xxxii=32
\mathchardef\@Mi=10001
\mathchardef\@Mii=10002
\mathchardef\@Miii=10003
\mathchardef\@Miv=10004

\newcount\@tempcnta
\newcount\@tempcntb
\newif\if@tempswa\@tempswatrue
\newdimen\@tempdima
\newdimen\@tempdimb
\newdimen\@tempdimc
\newbox\@tempboxa
\newskip\@tempskipa
\newskip\@tempskipb
\newtoks\@temptokena

\long\def\@whilenum#1\do #2{\ifnum #1\relax #2\relax\@iwhilenum{#1\relax
     #2\relax}\fi}
\long\def\@iwhilenum#1{\ifnum #1\expandafter\@iwhilenum
         \else\expandafter\@gobble\fi{#1}}
\long\def\@whiledim#1\do #2{\ifdim #1\relax#2\@iwhiledim{#1\relax#2}\fi}
\long\def\@iwhiledim#1{\ifdim #1\expandafter\@iwhiledim
        \else\expandafter\@gobble\fi{#1}}
\long\def\@whilesw#1\fi#2{#1#2\@iwhilesw{#1#2}\fi\fi}
\long\def\@iwhilesw#1\fi{#1\expandafter\@iwhilesw
         \else\@gobbletwo\fi{#1}\fi}
\def\@nnil{\@nil}
\def\@empty{}
\def\@fornoop#1\@@#2#3{}
\long\def\@for#1:=#2\do#3{%
  \expandafter\def\expandafter\@fortmp\expandafter{#2}%
  \ifx\@fortmp\@empty \else
    \expandafter\@forloop#2,\@nil,\@nil\@@#1{#3}\fi}
\long\def\@forloop#1,#2,#3\@@#4#5{\def#4{#1}\ifx #4\@nnil \else
       #5\def#4{#2}\ifx #4\@nnil \else#5\@iforloop #3\@@#4{#5}\fi\fi}
\long\def\@iforloop#1,#2\@@#3#4{\def#3{#1}\ifx #3\@nnil
       \expandafter\@fornoop \else
      #4\relax\expandafter\@iforloop\fi#2\@@#3{#4}}
\def\@tfor#1:={\@tf@r#1 }
\long\def\@tf@r#1#2\do#3{\def\@fortmp{#2}\ifx\@fortmp\space\else
    \@tforloop#2\@nil\@nil\@@#1{#3}\fi}
\long\def\@tforloop#1#2\@@#3#4{\def#3{#1}\ifx #3\@nnil
       \expandafter\@fornoop \else
      #4\relax\expandafter\@tforloop\fi#2\@@#3{#4}}
\long\def\@break@tfor#1\@@#2#3{\fi\fi}
\def\@removeelement#1#2#3{%
  \def\reserved@a##1,#1,##2\reserved@a{##1,##2\reserved@b}%
  \def\reserved@b##1,\reserved@b##2\reserved@b{%
    \ifx,##1\@empty\else##1\fi}%
  \edef#3{%
    \expandafter\reserved@b\reserved@a,#2,\reserved@b,#1,\reserved@a}}

\let\ExecuteOptions\@gobble

\def\on@line{ on input line \the\inputlineno}
\ifx\@ehc\@undefined\def\@ehc{}\fi

\def\@latex@error#1#2{\bgroup%
  \newlinechar`\^^J\def\MessageBreak{^^J\space\space#1: }%
  \edef\reserved@a{\egroup\errhelp{#2}\errmessage{#1}}%
  \reserved@a}

\bgroup\uccode`\!`\%\uppercase{\egroup
\def\@percentchar{!}}

\ifx\@@input\@undefined
 \let\@@input\input
\fi

\def\input{\@ifnextchar\bgroup\@iinput\@@input}
\def\@iinput#1{\input{#1} }

\ifx\filename@parse\@undefined
  \def\reserved@a{./}\ifx\@currdir\reserved@a
    \wlog{^^JDefining UNIX/DOS style filename parser.^^J}
    \def\filename@parse#1{%
      \let\filename@area\@empty
      \expandafter\filename@path#1/\\}
    \def\filename@path#1/#2\\{%
      \ifx\\#2\\%
         \def\reserved@a{\filename@simple#1.\\}%
      \else
         \edef\filename@area{\filename@area#1/}%
         \def\reserved@a{\filename@path#2\\}%
      \fi
      \reserved@a}
  \else\def\reserved@a{[]}\ifx\@currdir\reserved@a
    \wlog{^^JDefining VMS style filename parser.^^J}
    \def\filename@parse#1{%
      \let\filename@area\@empty
      \expandafter\filename@path#1]\\}
    \def\filename@path#1]#2\\{%
      \ifx\\#2\\%
         \def\reserved@a{\filename@simple#1.\\}%
      \else
         \edef\filename@area{\filename@area#1]}%
         \def\reserved@a{\filename@path#2\\}%
      \fi
      \reserved@a}
  \else\def\reserved@a{:}\ifx\@currdir\reserved@a
    \wlog{^^JDefining Mac style filename parser.^^J}
    \def\filename@parse#1{%
      \let\filename@area\@empty
      \expandafter\filename@path#1:\\}
    \def\filename@path#1:#2\\{%
      \ifx\\#2\\%
         \def\reserved@a{\filename@simple#1.\\}%
      \else
         \edef\filename@area{\filename@area#1:}%
         \def\reserved@a{\filename@path#2\\}%
      \fi
      \reserved@a}
  \else
    \wlog{^^JDefining generic filename parser.^^J}
    \def\filename@parse#1{%
      \let\filename@area\@empty
      \expandafter\filename@simple#1.\\}
  \fi\fi\fi
  \def\filename@simple#1.#2\\{%
    \ifx\\#2\\%
       \let\filename@ext\relax
    \else
       \edef\filename@ext{\filename@dot#2\\}%
    \fi
    \edef\filename@base{#1}}
  \def\filename@dot#1.\\{#1}
\else
  \wlog{^^J^^J%
    \noexpand\filename@parse was defined in texsys.cfg:^^J%
    \expandafter\strip@prefix\meaning\filename@parse.^^J%
    }
\fi

\long\def \IfFileExists#1#2#3{%
  \openin\@inputcheck#1 %
  \ifeof\@inputcheck
    \ifx\input@path\@undefined
      \def\reserved@a{#3}%
    \else
      \def\reserved@a{\@iffileonpath{#1}{#2}{#3}}%
    \fi
  \else
    \closein\@inputcheck
    \edef\@filef@und{#1 }%
    \def\reserved@a{#2}%
  \fi
  \reserved@a}
\long\def\@iffileonpath#1{%
  \let\reserved@a\@secondoftwo
  \expandafter\@tfor\expandafter\reserved@b\expandafter
             :\expandafter=\input@path\do{%
    \openin\@inputcheck\reserved@b#1 %
    \ifeof\@inputcheck\else
      \edef\@filef@und{\reserved@b#1 }%
      \let\reserved@a\@firstoftwo%
      \closein\@inputcheck
      \@break@tfor
    \fi}%
  \reserved@a}
\long\def \InputIfFileExists#1#2{%
  \IfFileExists{#1}%
    {#2\@addtofilelist{#1}\@@input \@filef@und}}

\chardef\@inputcheck0

\let\@addtofilelist \@gobble

\def\@defaultunits{\afterassignment\remove@to@nnil}
\def\remove@to@nnil#1\@nnil{}

\newdimen\leftmarginv
\newdimen\leftmarginvi

\newdimen\@ovxx
\newdimen\@ovyy
\newdimen\@ovdx
\newdimen\@ovdy
\newdimen\@ovro
\newdimen\@ovri
\newdimen\@xdim
\newdimen\@ydim
\newdimen\@linelen
\newdimen\@dashdim

\long\def\mbox#1{\leavevmode\hbox{#1}}

\let\@onlypreamble\@gobble

\let\protect\relax

\newdimen\fboxsep
\newdimen\fboxrule

\fboxsep = 3pt
\fboxrule = .4pt

\def\@height{height} \def\@depth{depth} \def\@width{width}
\def\@minus{minus}
\def\@plus{plus}
\def\hb@xt@{\hbox to}

\long\def\@begin@tempboxa#1#2{%
   \begingroup
     \setbox\@tempboxa#1{\color@begingroup#2\color@endgroup}%
     \def\width{\wd\@tempboxa}%
     \def\height{\ht\@tempboxa}%
     \def\depth{\dp\@tempboxa}%
     \let\totalheight\@ovri
     \totalheight\height
     \advance\totalheight\depth}
\let\@end@tempboxa\endgroup

\let\set@color\relax
\let\color@begingroup\relax
\let\color@endgroup\relax
\let\color@setgroup\relax

\let\color@hbox\relax
\let\color@vbox\relax
\let\color@endbox\relax


\begingroup
  \catcode`P=12
  \catcode`T=12
  \lowercase{
    \def\x{\def\rem@pt##1.##2PT{##1\ifnum##2>\z@.##2\fi}}}
  \expandafter\endgroup\x
\def\strip@pt{\expandafter\rem@pt\the}


\def\@input#1{%
  \IfFileExists{#1}{\@@input\@filef@und}{\message{No file #1.}}}

\def\@warning{\immediate\write16}

\input epsf.tex
\def\figin{\epsfcheck\figin}\def\figins{\epsfcheck\figins}
\def\epsfcheck{\ifx\epsfbox\UnDeFiSIeD
\message{(NO epsf.tex, FIGURES WILL BE IGNORED)}
\gdef\figin##1{\vskip2in}\gdef\figins##1{\hskip.5in}
\else\message{(FIGURES WILL BE INCLUDED)}%
\gdef\figin##1{##1}\gdef\figins##1{##1}\fi}
\def\DefWarn#1{}
\def\figinsert{\goodbreak\midinsert}
\def\ifig#1#2#3{\DefWarn#1\xdef#1{fig.~\the\figno}
\writedef{#1\leftbracket fig.\noexpand~\the\figno}%
\figinsert\figin{\centerline{#3}}\medskip\centerline{\vbox{\baselineskip12pt
\advance\hsize by -1truein\noindent\footnotefont{\bf Fig.~\the\figno:} #2}}
\bigskip\endinsert\global\advance\figno by1}

\lref\AldayKQ{
  L.~F.~Alday, J.~R.~David, E.~Gava and K.~S.~Narain,
  ``Towards a string bit formulation of N=4 super Yang-Mills,''
JHEP {\bf 0604}, 014 (2006).
[hep-th/0510264].
}

\lref\BerkovitsXU{
  N.~Berkovits,
  ``Quantum consistency of the superstring in AdS(5) x S**5 background,''
JHEP {\bf 0503}, 041 (2005).
[hep-th/0411170].
}

\lref\polyakov{
 A.~Polyakov, `'Old and New Aspects of the Strings/Gauge Correspondence,'' Strings 2002
proceedings, http://www.damtp.cam.ac.uk/strings02/avt/polyakov/.}

\lref\OoguriGX{
  H.~Ooguri and C.~Vafa,
  ``World sheet derivation of a large N duality,''
Nucl.\ Phys.\ B {\bf 641}, 3 (2002).
[hep-th/0205297].
}

\lref\GopakumarNS{
  R.~Gopakumar,
  ``From free fields to AdS,''
Phys.\ Rev.\ D {\bf 70}, 025009 (2004).
[hep-th/0308184].
}

\lref\Nastase{
  H.~Nastase,
  ``Towards deriving the AdS/CFT correspondence,''
[arXiv:1812.10347 [hep-th]].
}

\lref\GopakumarKI{
  R.~Gopakumar and C.~Vafa,
  ``On the gauge theory / geometry correspondence,''
Adv.\ Theor.\ Math.\ Phys.\  {\bf 3}, 1415 (1999), [AMS/IP Stud.\ Adv.\ Math.\  {\bf 23}, 45 (2001)].
[hep-th/9811131].
}

\lref\VerlindeIG{
  H.~L.~Verlinde,
  ``Bits, matrices and 1/N,''
JHEP {\bf 0312}, 052 (2003).
[hep-th/0206059].
}

\lref\BerensteinJQ{
  D.~E.~Berenstein, J.~M.~Maldacena and H.~S.~Nastase,
  ``Strings in flat space and pp waves from N=4 superYang-Mills,''
JHEP {\bf 0204}, 013 (2002).
[hep-th/0202021].
}

\lref\BerkovitsRJ{
  N.~Berkovits and C.~Vafa,
  ``Towards a Worldsheet Derivation of the Maldacena Conjecture,''
JHEP {\bf 0803}, 031 (2008), [AIP Conf.\ Proc.\  {\bf 1031}, 21 (2008)].
[arXiv:0711.1799 [hep-th]].
}

\lref\BerkovitsQC{
  N.~Berkovits,
  ``Perturbative Super-Yang-Mills from the Topological AdS(5) x S**5 Sigma Model,''
JHEP {\bf 0809}, 088 (2008).
[arXiv:0806.1960 [hep-th]].
}

\lref\BerkovitsYR{
  N.~Berkovits and O.~Chandia,
  ``Superstring vertex operators in an AdS(5) x S**5 background'',
Nucl.\ Phys.\ B {\bf 596}, 185 (2001).
[hep-th/0009168].
}

\lref\BargheerNNE{
  T.~Bargheer, J.~Caetano, T.~Fleury, S.~Komatsu and P.~Vieira,
  ``Handling Handles: Nonplanar Integrability in $N=4$ Supersymmetric Yang-Mills Theory,''
Phys.\ Rev.\ Lett.\  {\bf 121}, no. 23, 231602 (2018).
[arXiv:1711.05326 [hep-th]].
}

\lref\GaiottoYB{
  D.~Gaiotto and L.~Rastelli,
  ``A Paradigm of open / closed duality: Liouville D-branes and the Kontsevich model,''
JHEP {\bf 0507}, 053 (2005).
[hep-th/0312196].
}

\lref\BerkovitsRB{
  N.~Berkovits,
``Covariant quantization of the superparticle using pure spinors,''
JHEP {\bf 0109}, 016 (2001).
[hep-th/0105050].
}

\lref\GalperinAV{
  A.~Galperin, E.~Ivanov, S.~Kalitsyn, V.~Ogievetsky and E.~Sokatchev,
``Unconstrained N=2 Matter, Yang-Mills and Supergravity Theories in Harmonic Superspace,''
Class.\ Quant.\ Grav.\  {\bf 1}, 469 (1984)..
}

\lref\vallilo{
  B.~Vallilo and L.~Mazzucato,
  ``The Konishi Multilpet at Strong Coupling,''
JHEP {\bf 1112}, 029 (2011).
[arXiv:1102.1219 [hep-th]].
}

\lref\MikhailovAF{
  A.~Mikhailov,
 ``Finite dimensional vertex,''
JHEP {\bf 1112}, 005 (2011).
[arXiv:1105.2231 [hep-th]].
}

\lref\mikh{
  A.~Mikhailov and R.~Xu, ``BRST cohomology of the sum of two pure spinors,''
to appear.
}

\lref\minahan{
  J.~Minahan,
  ``Holographic three-point functions for short operators,''
JHEP {\bf 1207}, 187 (2012).
[arXiv:1206.3129 [hep-th]].
}

\lref\BerkovitsGA{
  N.~Berkovits,
  ``Simplifying and Extending the AdS(5) x S**5 Pure Spinor Formalism,''
JHEP {\bf 0909}, 051 (2009).
[arXiv:0812.5074 [hep-th]].}

\lref\BerkovitsPX{
  N.~Berkovits,
  ``Multiloop amplitudes and vanishing theorems using the pure spinor formalism for the superstring,''
JHEP {\bf 0409}, 047 (2004).
[hep-th/0406055].
}



\lref\BerkovitsBT{
  N.~Berkovits,
  ``Pure spinor formalism as an N=2 topological string'',
JHEP {\bf 0510}, 089 (2005).
[hep-th/0509120].}



\lref\BerkovitsFE{
  N.~Berkovits,
  ``Super Poincare covariant quantization of the superstring'',
JHEP {\bf 0004}, 018 (2000).
[hep-th/0001035].
}


\lref\MazzucatoJT{
  L.~Mazzucato,
  ``Superstrings in AdS'',
[arXiv:1104.2604 [hep-th]].
}


\lref\MikhailovMDO{
  A.~Mikhailov,
  ``A minimalistic pure spinor sigma-model in AdS,''
JHEP {\bf 1807}, 155 (2018).
[arXiv:1706.08158 [hep-th]].
}

\lref\HeslopNP{
  P.~Heslop and P.~S.~Howe,
  ``Chiral superfields in IIB supergravity'',
Phys.\ Lett.\ B {\bf 502}, 259 (2001).
[hep-th/0008047].
}







\lref\AzevedoRVA{
  T.~Azevedo and N.~Berkovits,
  ``Open-closed superstring amplitudes using vertex operators in ${AdS}_5 \times S^5$,''
JHEP {\bf 1502}, 107 (2015).
[arXiv:1412.5921 [hep-th]].
}

\lref\SohniusWK{
  M.~F.~Sohnius,
  ``Bianchi Identities for Supersymmetric Gauge Theories,''
Nucl.\ Phys.\ B {\bf 136}, 461 (1978).
}

\lref\ArutyunovGA{
  G.~Arutyunov and S.~Frolov,
  ``Foundations of the $AdS_5 \, x \, S^5$ Superstring. Part I,''
J.\ Phys.\ A {\bf 42}, 254003 (2009).
[arXiv:0901.4937 [hep-th]].
}

\lref\mt{
  R.~R.~Metsaev and A.~A.~Tseytlin,
  ``Type IIB superstring action in AdS(5) x S**5 background,''
Nucl.\ Phys.\ B {\bf 533}, 109 (1998).
[hep-th/9805028].
}



\lref\HoweSRA{
  P.~S.~Howe and P.~C.~West,
  ``The Complete N=2, D=10 Supergravity'',
Nucl.\ Phys.\ B {\bf 238}, 181 (1984).
}

\lref\deWit{
  B.~de Wit, J.~Hoppe and H.~Nicolai,
  ``On the Quantum Mechanics of Supermembranes,''
Nucl.\ Phys.\ B {\bf 305}, 545 (1988)..
}

\lref\WittenNN{
  E.~Witten,
  ``Perturbative gauge theory as a string theory in twistor space,''
Commun.\ Math.\ Phys.\  {\bf 252}, 189 (2004).
[hep-th/0312171].
}

\lref\WittenXJ{
  E.~Witten,
  ``Topological Sigma Models,''
Commun.\ Math.\ Phys.\  {\bf 118}, 411 (1988)..
}


\lref\BonelliRV{
  G.~Bonelli and H.~Safaai,
  ``On gauge/string correspondence and mirror symmetry,''
JHEP {\bf 0806}, 050 (2008).
[arXiv:0804.2629 [hep-th]].
}








\def\bar{\overline}

\def\a{{\alpha}}

\def\l{{\lambda}}

\def\b{{\beta}}

\def\g{{\gamma}}

\def\d{{\delta}}
\def\e{{\epsilon}}
\def\s{{\sigma}}

\def\half{{1\over 2}}
\def\p{{\partial}}

\def\t{{\theta}}

\def\th{{\widehat\theta}}

\Title{\vbox{\baselineskip12pt
\hbox{}}}
{{\vbox{\centerline{Sketching a Proof of the }
\smallskip
\centerline{Maldacena Conjecture at Small Radius}}} }
\bigskip\centerline{Nathan Berkovits\foot{e-mail: nathan.berkovits@unesp.br}}
\bigskip
\centerline{\it ICTP South American Institute for Fundamental Research}
\centerline{\it Instituto de F\'\i sica Te\'orica, UNESP - Univ. 
Estadual Paulista }
\centerline{\it Rua Dr. Bento T. Ferraz 271, 01140-070, S\~ao Paulo, SP, Brasil}
\bigskip

\vskip .3in

At small AdS radius, the superstring on $AdS_5\times S^5$ was conjectured by Maldacena to be equivalent to ${\cal N}=4$ super-Yang-Mills at small `t Hooft coupling where thickened Feynman diagrams can be used to compute scattering amplitudes. It was previously shown that the pure spinor worldsheet action of the $AdS_5\times S^5$ superstring can be expressed as the sum of a BRST-trivial term and a ``B-term'' which is antisymmetric in worldsheet derivatives. Using the explicit form of the pure spinor vertex operators, it will be argued here that the free super-Yang-Mills Feynman diagrams are described by the BRST-trivial term where the thickened propagators are the regions of the string worldsheet near the AdS boundary and the holes are the regions near the AdS horizon. Evidence will then be presented that the antisymmetric B-term generates the super-Yang-Mills vertex so that, at small radius and arbitrary genus, the superstring amplitudes correctly reproduce the super-Yang-Mills Feynman diagram expansion. 
\vskip .3in

\Date {March 2019}
\newsec{Introduction}

Although string theory in an $AdS_5\times S^5$ background has mostly been studied at large AdS radius where the supergravity approximation is
valid, there have been several approaches \GopakumarKI\polyakov\BerensteinJQ\OoguriGX\VerlindeIG\GopakumarNS\WittenNN\GaiottoYB\AldayKQ\BerkovitsRJ\BonelliRV\BerkovitsQC\BargheerNNE\Nastase\ to studying $AdS_5\times S^5$ string theory at small radius where the dual super-Yang-Mills theory is weakly coupled. Using the pure spinor formalism, BRST invariance and $PSU(2,2|4)$ invariance of this background imply that the worldsheet action is not
renormalized and can be expressed for arbitrary radius $R$ as \BerkovitsFE\BerkovitsXU
\eqn\actionpure{ S = R^2 \int d^2 z ( \half J_2 \bar J_2  - {3\over 4} J_1 \bar J_3 - {1\over 4}J_3 \bar J_1) + {\rm ghost~~ terms}}
where the ghost terms describe the coupling of the pure spinor worldsheet ghosts and $(J_1, J_2, J_3)=g^{-1} \p g$ and $(\bar J_1, \bar J_2, \bar J_3)=g^{-1}\bar\p g$ are the usual left-invariant currents \mt\ constructed
from the supercoset $g\in {{PSU(2,2|4)}\over{SO(4,1)\times SO(5)}}$ that parametrizes $AdS_5\times S^5$.

As was shown in \BerkovitsGA, this worldsheet action can expressed as the sum of a BRST-trival term and an antisymmetric $B$-term as
\eqn\actionexp{S = Q\Lambda + R^2 \int d^2 z B \quad{\rm where}}
\eqn\bfield{B = {1\over 4} (J_3 \bar J_1 - \bar J_3  J_1) + ...}
and $...$ includes terms depending on the pure spinor ghosts.
So at small radius, the string theory can be studied by expanding around the BRST-trival term with the perturbation $R^2 \int d^2 z B.$
It will be argued here that this expansion reproduces the standard Feynman diagram expansion of super-Yang-Mills at small `t Hooft coupling $\l_{tHooft}$, thereby proving the Maldacena conjecture at small radius.

Using the explicit form of the pure spinor vertex operators, it will first be argued that the topological string described by \actionexp\ at $R=0$ reproduces the free
super-Yang-Mills Feynman diagrams where the Feynman propagators are the regions of the string worldsheet near the AdS boundary and the holes in the
Feynman diagram are the regions of the string worldsheet near the AdS horizon. These holes will be related to $D_3$ branes in a manner similar to the closed-open string dualities discussed in \GopakumarKI\OoguriGX\BerkovitsRJ.

Evidence will then be presented that the $B$-term in the action of \actionexp\ generates the cubic super-Yang-Mills vertex proportional to $R^2 \sim \sqrt{\l_{tHooft}}$ where antisymmetric terms of the type
$f (\p g \bar \p h - \bar\p g \p h)$ in $B$ generate commutator terms of the type $f [g,h]$ in the cubic vertex. And since the genus $g$ string amplitude is proportional to $(g_s)^{2g-2}  \sim  N^{2-2g}$, one obtains the usual `t Hooft expansion in $1\over N$ for the non-planar Feynman diagrams.

Section 2 of this paper will describe the structure of $AdS_5\times S^5$ vertex operators at small radius, section 3 will discuss the relation of the topological action and free super-Yang-Mills, section 4 will compare the $B$ term and the cubic super-Yang-Mills vertex, and the Appendix will review the construction of the $AdS_5\times S^5$ topological action.

\newsec{AdS Vertex Operators at Small Radius}

\subsec{Half-BPS vertex operators}

To construct $AdS_5\times S^5$ vertex operators at small AdS radius, it will be useful to first consider the half-BPS vertex operators describing supergravity states, since their form is expected to be independent of the radius. As in any supergravity background, these vertex operators can be expressed as \BerkovitsYR
\eqn\any{
V = \l_L^\a \l_R^\b A_{\a\b} (X, Y, \t_L,\t_R)}
satisfying $QV =0$
where $X\in AdS_5$, $Y\in S^5$, $(\t_L^\a, \t_R^\a)$ are the fermionic variables for $\a=1$ to 16, $\l_L^\a$ and $\l_R^\a$ are pure spinors satisfying
$\l_L\g^a\l_L = \l_R\g^a\l_R=0$ for $a=0$ to 9, and $Q = \l_L^\a \nabla_{L\a} +
\l_R^\a \nabla_{R\a}$ is the pure spinor
BRST operator. 

To express these vertex operators in $PSU(2,2|4)$ covariant form, parameterize the ${{PSU(2,2|4)}\over{SO(4,1) \times SO(5)}}$ coset as
\eqn\paramone{ g(X,Y, \t, \th) = \exp (\t^R_J q^J_R) \exp (\t^J_R  q_J^R) G^R_{\tilde R}(X) H^J_{\tilde J}(Y)}
where $G^R_{\tilde R}(X)$ is an ${{SO(4,2)}\over{SO(4,1)}}$ coset for $AdS_5$, $H^J_{\tilde J}(Y)$ is an ${{SO(6)}\over{SO(5)}}$ coset for $S^5$,
$R=1$ to 4 and $J=1$ to 4 are $SU(2,2)$ and $SU(4)$ spinor indices, 
$\tilde R=1$ to 4 and $\tilde J=1$ to 4 are $SO(4,1)$ and $SO(5)$ spinor indices, $(q^J_R,  q_J^R)$ are the 32 fermionic generators of $PSU(2,2|4)$, and $X^{RS}$ and $Y^{JK}$ are $SO(4,2)$ and $SO(6)$ vectors normalized to satisfy
$\e_{RSTU} X^{RS} X^{TU} = \e_{JKLM} Y^{JK} Y^{LM} =8$. Under global $PSU(2,2|4)$ transformations parameterized by $\Sigma$, 
\eqn\glob{\d g = \Sigma g  + g \Omega} 
where $\Omega$ is a local $SO(4,1)\times SO(5)$ gauge transformation. And under a BRST transformation,
 \eqn\brst{\d g = g [(\l_L +  \l_R)^{\tilde R}_{\tilde J} q^{\tilde J}_{\tilde R} + 
 ( \l_L -  \l_R)^{\tilde J}_{\tilde R}  q^{\tilde R}_{\tilde J} ] + g \Omega'}
where $\Omega'$ is another local $SO(4,1)\times SO(5)$ gauge transformation. Under these gauge transformations,
$(\l_L, \l_R)$ transform as $SO(4,1)\times SO(5)$ spinors,
i.e. 
\eqn\latransf{\d (\l_L)_ {\tilde J}^{\tilde R} = (\Omega +\Omega')^{\tilde R}_{\tilde S} (\l_L)_{ \tilde J}^{\tilde S} -
 (\Omega +\Omega')^{\tilde K}_{\tilde J} (\l_L)_{ \tilde K}^{\tilde R} , }
 $$
 \d (\l_R)_{ \tilde J}^{\tilde R} = (\Omega +\Omega')^{\tilde R}_{\tilde S} (\l_R)_{ \tilde J}^{\tilde S} -
 (\Omega +\Omega')^{\tilde K}_{\tilde J} (\l_R)_{ \tilde K}^{\tilde R} .$$

Since the vertex operators must be invariant under the local $SO(4,1)\times SO(5)$ gauge transformations, it is convenient to define
$SO(4,1)\times SO(5)$ gauge-invariant worldsheet ghost variables as
\eqn\ghost{(\tilde\l_L)^R_J = G^R_{\tilde R} (H^{-1})_J^{\tilde J} (\l_L)^{\tilde R}_{\tilde J}, \quad
(\tilde\l_R)^R_J = G^R_{\tilde R} (H^{-1})_J^{\tilde J} (\l_R)^{\tilde R}_{\tilde J}}
$$(\tilde\l_L)^J_R = (G^{-1})_R^{\tilde R} H^J_{\tilde J} ( \l_L)_{\tilde R}^{\tilde J}, \quad
(\tilde\l_R)^R_J = (G^{-1})_R^{\tilde R} H^J_{\tilde J} ( \l_R)_{\tilde R}^{\tilde J}$$
which transform as $SO(4,2)\times SO(6)$ spinors under the global isometries. Note that
\eqn\lambdaeq{(\tilde\l_L)^R_J = X^{RS} Y_{JK} (\tilde\l_L)^K_S, \quad (\tilde\l_R)^R_J = - X^{RS} Y_{JK} (\tilde\l_R)^K_S}
where
the $AdS_5$ and $S^5$ variables $X^{RS}$ and $Y^{JK}$ are defined in terms of the parameterization of \paramone\ as
\eqn\defxy{ X^{RS} = G^R_{\tilde R} \sigma_6^{\tilde R\tilde S} G^S_{\tilde S}, \quad
Y^{JK} = H^J_{\tilde J} \sigma_6^{\tilde J\tilde K} H^K_{\tilde K}}
and $\sigma_6$ is the $4\times 4$ matrix which commutes with the $SO(4,1)$ and $SO(5)$ Pauli matrices.
The pure spinor condition $ \l_L\g^a\l_L = \l_R \g^a \l_R =0$ for $a=0$ to 9 implies
that $\tilde\l_L$ and $\tilde\l_R$ satisfy
\eqn\purel{(\tilde\l_L)^R_J (\tilde\l_L)^J_S ={1\over 4} \d^R_S (\tilde\l_L)^2, \quad (\tilde\l_L)^J_R (\tilde\l_L)^R_K = {1\over 4}\d^J_K (\tilde\l_L)^2, }
$$(\tilde\l_L)^{[J}_R (\tilde\l_L)^{K]}_S ={1\over 2} \e^{JKLM} \e_{RSTU} (\tilde\l_L)_L^T (\tilde\l_L)_M^U,$$
and similarly for $\tilde\l_R$.

In terms of these $SO(4,1)\times SO(5)$ gauge-invariant variables, one can easily construct the half-BPS vertex operators in a
$PSU(2,2|4)$-invariant manner.
First consider the supergravity state dual to the super-Yang-Mills state $\Tr((\Phi_{12}(0))^n)$ where $\Phi_{JK}(x)$ are the six super-Yang-Mills scalars and $\Phi_{12}(0)$ is the complex scalar located at $x^m=0$ with charge $+1$ with respect to a U(1) subgroup $J$ of the $SO(6)$ R-symmetries. This state is described by the BRST-invariant vertex operator
\eqn\vsimple{V = (\l_L \l_R)  ({{Y_{12}}\over {X_{12}}})^n \prod_{A=1}^8 \t^A \d (Q(\t^A)) }
where $X_{12}$ carries $+  1$ dilatation charge, $Y_{12}$ carries $+1$ U(1) charge, $\t^A$ for $A=1$ to 8 are the fermionic variables which carry $+\half$ dilatation charge and $+\half$ U(1) charge, and $(\l_L\l_R) \equiv \l_L^\a \g^{01234}_{\a\b} \l_R^\b$ is the unintegrated vertex operator of ghost-number 2 for the radius modulus.

To verify that \vsimple\ is BRST-invariant, first note that $(\l_L \l_R)$ and $\prod_{A=1}^8 \t^A \d (Q(\t^A))$ are BRST invariant where the operator $\t^A \d (Q(\t^A))$
has the form of a picture-lowering operator as in \BerkovitsPX. If $(\Delta + J)$ charge is the sum of the dilatation and U(1) charge, the anticommutation of $\{q,q\}$ only generates a transformation of ${{Y_{12}}\over{X_{12}}}$ when one of the $q$'s carry $+1$ $(\Delta +J)$ charge and the other $q$ carries $-1$ $(\Delta + J)$ charge. As will now be shown, this implies that 
${{Y_{12}}\over{X_{12}}}$ is BRST invariant when multiplied by $\prod_{A=1}^8\t^A \d(Q(\t^A))$. 

When the $q$ carrying $-1$ $(\Delta + J)$ charge comes from the BRST operator, the BRST transformation of $g$ in \brst\ is proportional to $Q(\t^A)$. And when the $q$
carrying $-1$ $(\Delta + J)$ charge comes from $g$, the BRST transformation of $g$ in \brst\ is proportional to $\t^A$.  So in both cases, the BRST transformation is cancelled
by the factor of  $\prod_{A=1}^8\t^A \d(Q(\t^A))$. For a similar reason, \vsimple\ is invariant under all supersymmetries except for the 8 $q$'s which carry $-1$ $(\Delta + J)$ charge, which are the same 24 supersymmetries that leave invariant the scalar $\Phi_{12}(0)$.

To relate $V$ of \vsimple\ with the usual unintegrated vertex operator of \any, one needs to hit $V$ with the eight picture-raising operators $Q(\xi_A)$ where, using Friedan-Martinec-Shenker bosonization, 
$\tilde\l^A = \eta^A e^{\phi_A}$ for $A =1$ to 8 are the eight components of $\tilde\l^\a$ with $+1$ $(\Delta + J)$ charge and $\xi_A$ are the conjugate momenta to $\eta^A$. More explicitly, one can use the relation $\xi_1 \d (\tilde\l^1) = \xi_1 e^{-\phi^1} = (\tilde\l^1)^{-1}$ to write
\eqn\raising{ Q(\xi_1) V = [ Q, (\l_L\l_R) ({{Y_{12}}\over {X_{12}}})^n {{\t^1}\over{\tilde\l^1}} \t^2 \d(\tilde\l^2) ... \t^8 \d(\tilde\l^8) ].}
Note that $QV=0$ implies that $[ Q, (\l_L\l_R) ({{Y_{12}}\over {X_{12}}})^n \t^1 \t^2 \d(\tilde\l^2) ... \t^8 \d(\tilde\l^8) ]$ is proportional to $\tilde\l^1$, so
 \raising\ has no poles when $\tilde\l^1 =0$. As will be shown in a future paper, the vertex operator obtained after hitting \vsimple\ with eight picture-raising operators has the expected form of \any\ for the half-BPS state dual to $\Tr((\Phi_{12}(0))^n)$. 

\subsec { General non-BPS vertex operators at small radius}

For the half-BPS state dual to $\Tr((\Phi_{12}(0))^n)$, the unintegrated vertex operator of \vsimple\ can be expressed in the ``zero picture" as
\eqn\vdouble{ V = (\l_L \l_R) ~C~D~ ({{Y_{12}}\over {X_{12}}})^n }
where $C= \prod_{A=1}^8 Q(\xi_A)$ is the "picture-raising" operator and $D = \prod_{A=1}^8\t^A \d(Q(\t^A))$ is the ``picture-lowering" operator. Since adding an equal number of picture-raising and picture-lowering operators is a BRST-trivial operation, one can also write
\eqn\vtriple{ V = (\l_L \l_R) (C~D~ {{Y_{12}}\over {X_{12}}})^n .}
And all other half-BPS vertex operators can be obtained by hitting \vtriple\ with the appropriate $PSU(2,2|4)$ generators.

At zero radius, the closed string states can be represented as ``necklaces" made of ``beads" where each bead is a free super-Yang-Mills state. This suggests writing the half-BPS vertex operator as
\eqn\vcomp{ V = (\l_L \l_R (0))  }
$$C(\s_1 - \e)~D(\s_1) {{Y_{12}}\over {X_{12}}}(\s_1) 
C(\s_2 -\e) ~ D(\s_2) {{Y_{12}}\over {X_{12}}}(\s_2) ...
 C(\s_n -\e) ~ D(\s_n) {{Y_{12}}\over {X_{12}}}(\s_n) $$
 where $(\s_1, ..., \s_n)$ are $n$ cyclically ordered points on a small closed string which mark the locations of the``beads", the picture-raising operators $C$ are placed between the beads on the necklace, and the operator $(\l_L\l_R)$ is placed at the center of the small necklace. Since the operators $(\l_L\l_R)$, $C$ and $D{{Y_{12}}\over {X_{12}}}$ are all BRST-closed, $QV=0$. And by hitting $D{{Y_{12}}\over {X_{12}}}$ with different $PSU(2,2|4)$ generators at the different beads, $V$ can be easily generalized for an arbitrary non-BPS state at zero radius to the vertex operator
 \eqn\varb{ V = (\l_L \l_R (0))  C(\s_1 - \e)~E_1 (\s_1) ~C(\s_2 -\e) ~ E_2(\s_2) ...
 C(\s_n -\e) ~ E_n(\s_n)}
 where $E(\s)$ is obtained from $D(\s){{Y_{12}}\over {X_{12}}}(\s)$ by acting with the $PSU(2,2|4)$ transformation which takes $\Phi_{12}(0)$ into the desired super-Yang-Mills state. Note that for half-BPS states, the cyclic ordering of the $E$'s in \varb\ is irrelevant since the $E$'s have no singular OPE's with each other. But for non-BPS states, the $E$'s have 
singular OPE's with each other so normal-ordering of \varb\ needs to be performed, and different cyclic orderings of the $E$'s describe different vertex operators. 

\newsec{Free Super-Yang-Mills}

\subsec{Topological action}

As reviewed in the appendix, it was shown in \BerkovitsGA\ that if one assumes $(\l_L \l_R)$ is non-vanishing so that $(\l_L \l_R)^{-1}$ is well-defined, the pure spinor $AdS_5\times S^5$ superstring worldsheet action can be expressed as 
\eqn\actionexp{S = Q\Lambda + R^2 \int d^2 z B \quad{\rm where}}
\eqn\bfield{B = {1\over 4} (J_3 \eta \bar J_1 - \bar J_3 \eta  J_1) - {{(\l_L\eta  \g_{ab}\l_R) + (\l_L\g_{ab}\eta\l_R)} \over{4(\l_L \l_R)}}  J_2^{a} \bar J_2^{b} }
$$- {{(\l_L\eta \g_a\eta \bar J_1) (\l_L \g^a J_1)}\over{4(\l_L \l_R)}} + {{(\l_R\eta \g_a \eta J_3) (\l_R \g^a\bar  J_3)}\over{4(\l_L \l_R)}}  $$
and $\eta_{\a\b} \equiv \g^{01234}_{\a\b}$.
At zero radius, the worldsheet action of \actionexp\ becomes BRST-trivial and the $n$-point genus $g$ scattering amplitude ${\cal A}_{n,g}$ reduces to an integral over the worldsheet zero modes of $(x,\t,\l)$ of the vertex operator insertions, i.e.
\eqn\ampfree{{\cal A}_{n,g} = \langle V_1 (z_1) ... V_n (z_n)\rangle_g}
where $(z_1, ..., z_n)$ are arbitrary points on the genus $g$ worldsheet and sufficient powers of $(\l_L \l_R)^{-1}$ are included in the vertex operators of \varb\ so that ${\cal A}_{n,g}$ has the appropriate ghost number to be non-zero at genus $g$. Although one naively might think one needs to integrate the vertex operator locations $z_r$ over the worldsheet and integrate the parameters of the genus $g$ worldsheet over Teichmuller moduli space, these integrals are unnecessary since the worldsheet action is independent
of the worldsheet metric \WittenXJ. So the amplitudes are ``topological", i.e. are independent of the choice of $z_r$ and Teichmuller parameters. It will now be argued that \ampfree\ correctly reproduces the correlation functions of free super-Yang-Mills computed using the thickened Feynman diagrams.

\subsec{Emergence of propagators}

The first step will be to argue that the worldsheet splits into regions which are close to the AdS boundary and regions which are close to the AdS horizon.
At the locations of the picture-lowering operators $D$, the components of the bosonic worldsheet ghosts $(\tilde\l_L^A,\tilde\l_R^A)$ with $+1$ $(\Delta +J)$ charge vanish. And at the locations of the picture-raising operators $C$, these same ghost components diverge. Note that the relation of \ghost\ implies that $\tilde \l_L^A \sim \sqrt{ z_{AdS}} \l_L^A$ and $\tilde \l_R^A \sim \sqrt{z_{AdS}}\l_R^A$ where $z_{AdS}$ is the fifth $AdS_5$ coordinate which measures the distance to the boundary. So if the original $SO(4,1)\times SO(5)$ pure spinor ghosts $(\l_L, \l_R)$ are regular at the locations of these operators, the relation \ghost\ implies that $z_{AdS}$ is near the boundary (i.e. $z_{AdS}\to 0$) at the picture-lowering operator locations and is near the horizon (i.e. $z_{AdS}\to \infty$) at the picture-raising operator locations. 

When $z_{AdS}$ is finite, the exponential of the topological worldsheet action is
\eqn\topz{ \exp ( - \Lambda \int d\tau d\s [ z_{AdS}^{-2} \p x^m \bar\p x_m + ...]).}
Since the action is BRST-trivial, one can take $\Lambda\to\infty$ so that all non-zero modes of the worldsheet variables must vanish and only the constant worldsheet modes contribute to the functional integral. However, when $z_{AdS}\to \infty$, the inverse factor of $z_{AdS}$ in the worldsheet action means that the four $x^m$ variables of $+1$ dilatation charge and the sixteen $\t$ variables with $+\half$ dilatation charge can be discontinuous. So the worldsheet splits into regions separated by the $z_{AdS}=\infty$ discontinuity where the $(x^m, \t^\a)$ zero modes can take different values in the disconnected regions. However, the five $S^5$ variables of zero dilatation charge and the 16 $\t^\a$ variables with $-\half$ dilatation charge have no discontinuities at $z_{AdS}\to \infty$, so they take the same value in all regions. Therefore, the discontinuities that separate the different regions are similar to $D_3$-branes located at $z_{AdS}=\infty$ and a fixed point $Y= y_0$ of $S^5$. 

Each of the disconnected regions contains at least one ``bead" which is located at $z_{AdS}=0$, so these regions are all near the AdS boundary. Suppose one of the regions contains $r$ beads, so that its contribution to the amplitude is proportional to 
\eqn\propbead{
 \int d^4 x \int d^{11} \l \int d^{16}\t~E_1 E_2 ... E_r |_{z_{AdS}=0, Y = y_0}.}
Since each $E$ contributes 8 $\t$'s and there are 16 $\t$ zero modes, one
easily sees that \propbead\ vanishes unless $r=2$. So each disconnected region must contain precisely two beads. Therefore, the worldsheet splits into ``thickened propagators" near the AdS boundary which
connect two beads, and which are separated by ``$D_3$ branes" located at $z_{AdS}=\infty$ and $Y=y_0$ that connect picture-raising operators. For example, see Figure 1 at the end of this paper for a worldsheet which splits into three thickened propagators near the AdS boundary and two $D_3$-brane holes near the AdS horizon.

Furthermore, one can
argue by $PSU(2,2|4)$ symmetry that the contribution of \propbead\ when $r=2$ is proportional to the standard propagator for the super-Yang-Mills states described by $E_1$ and $E_2$. For example, if $E_1$ and $E_2$ correspond to Yang-Mills scalars $\Phi^{JK}(x_1)$ and $\Phi^{LM}(x_2)$ as in \vsimple, 
\eqn\canuse{\int d^4 x \int d^{11}\l\int d^{16}\t  ~E^{JK}_1(x_1) E^{LM}_2(x_2) |_{z_{AdS}=0, Y = y_0}}
$$ \sim   
\e^{JKLM} \lim_{z_{AdS}\to 0} \int d^4 x \int d^{11} \l f(x, \l)  {z_{AdS}\over{(x-x_1)^2 + z_{AdS}^2}} {z_{AdS}\over{(x-x_2)^2 + z_{AdS}^2}}$$
where the factor of $\e^{JKLM}$ comes from integration over the 16 $\t$'s in $E_1 E_2$, and the factor of $f(x,\l)$ comes from writing the 16 $\d (\tilde\l)$ factors in $E_1 E_2$ in terms of the $(x^m, \l^\a)$ coordinates. Note that only the points $x^m=x^m_1$ and $x^m=x_2^m$ contribute to \canuse\ in the limit $z_{AdS}\to 0$, and assuming that the factor of $f(x,\lambda)$ cancels the integration over $d^4 x$ and $d^{11}\l$, \canuse\ reproduces the expected propagator
 ${{\e^{JKLM}} \over {(x_1 - x_2)^2}}$.
It would be interesting to
better understand how to integrate over the $\tilde\l$ variables in this topological string and compute the factor of $f(x,\l)$.

\subsec{Topological amplitudes}

For the scattering amplitude defined in \ampfree, the worldsheet splits into propagators and holes so that ${\cal A}_{n,g}$ reproduces the standard
computation for super-Yang-Mills using thickened Feynman diagrams in the absence of vertices. For example, Figure 1 describes the worldsheet of a three-point
tree-level amplitude ${\cal A}_{3,0}$ where the three vertex operators correspond to $Tr(\Phi^2)$ super-Yang-Mills operators. 
 
Furthermore, on a worldsheet of genus $g$, there is the standard factor of $(g_S)^{2g-2}$ where $g_s = {{\lambda_{tHooft}}\over N}$ is the string coupling constant and $N$ is the number of colors. Using the 't Hooft expansion at large $N$, thickened Feynman diagrams of genus $g$ carry a factor of $N^{2-2g} = \lambda_{tHooft}^{2-2g} (g_s)^{2g-2}$.
So up to a factor of $\lambda_{tHooft}^{2g-2}$, the genus $g$ string scattering amplitude at zero radius correctly reproduces the standard Feynman diagram rules for free super-Yang-Mills with gauge group $SU(N)$.

One can also use this topological string to define the closed-open vertex by computing the disk amplitude of one closed string state and $n$ open string
massless super-Yang-Mills states. Note that a similar closed-open vertex was defined in \AzevedoRVA\ for half-BPS states, which will now be generalized at zero `t Hooft coupling to arbirary closed string states. 
The closed-open vertex should vanish unless the closed string state is dual to the $n$ super-Yang-Mills states, i.e. unless the states $E_1 (\s_1)$ ... $E_n(\s_n)$
in \varb\ coincide with the cyclically ordered $n$ super-Yang-Mills states described by the open string vertex operators. 

For the worldsheet of this disk amplitude, the ends of the
open strings are located at $n$ $D_3$-branes near the AdS horizon which are connected to the ``holes" containing the $n$ picture-raising operators $C(\s)$ in the closed string 
vertex operator of \varb. For example, see Figure 2 at the end of this paper for a disk amplitude with three open string vertex operators.

\newsec{Cubic Super-Yang-Mills Interaction}

\subsec{Commutators from B terms}

Since the BRST-trivial term in the worldsheet action generates the free super-Yang-Mills diagrams and the complete $AdS_5\times S^5$ worldsheet action is 
\eqn\comp{S = Q\Omega + R^2 \int d\tau d\s B,}
it is natural to conjecture that $\int d\tau d\s B$ generates the cubic vertex in the super-Yang-Mills Feynman diagrams
where $R^2 = \sqrt{g_{YM}^2 N} = \sqrt{\l_{tHooft}}$,
\eqn\Bdef{B = {1\over 4}\eta_{\a\b} J_{1}^\a \wedge J_{3}^\b   - {{(\l_L\eta \g_{ab}\l_R) + (\l_L \g_{ab}\eta\l_R)}\over{8(\l_L\l_R)}} J_{2}^{a} \wedge J_{2}^{b} }
$$ +
{{(\l_L\g^a J_{1})\wedge (\l_L\eta \g_a\eta J_{1})}\over {8(\l_L\l_R)}} +{{(\l_R\g^a J_{3})\wedge (\l_R\eta \g_a\eta J_{3})}\over {8(\l_L\l_R)}} ,$$
and $J^A\wedge J^B \equiv J_\sigma^A J_\tau^B - J_\tau^A J_\sigma^B$.
Note that by introducing an auxiliary
field $D_{ab}$ for $a,b=0$ to 9, the complete super-Yang-Mills action can be expressed as the sum of the quadratic and cubic terms
\eqn\sym{S_{YM} = \int d^4 x \Tr (D_{ab} D^{ab} + \l_{tHooft}  \psi^\a \g^a_{\a\b} [A_a, \psi^\b] + \l_{tHooft} D^{ab} [A_a, A_b] )}
where $A_a = {1\over \l_{tHooft}} \p_a + {\cal A}_a$ for $a=0$ to 3 are the four covariant spacetime derivatives and $A_a = \Phi_{JK}$ for $a=4$ to 9 are the six scalars.

To verify this conjecture, first divide the worldsheet integration of $B$ into small squares of length $\Delta \tau$ and height $\Delta \s$. Each term in $B$ has the form
\eqn\formb{\int d\tau \int d\s [f( \p_\tau g \p_\s h - \p_ \s g \p_\tau h)] = }
$$\Delta \tau \Delta\sigma  f(\tau,\s) [
{{g(\tau + \Delta\tau, \s) - g(\tau,\s)}\over{\Delta\tau}}
{{h(\tau, \s + \Delta\s) - h(\tau,\s)}\over{\Delta\s}}$$
$$ - 
{{g(\tau, \s + \Delta\s) - g(\tau,\s)}\over{\Delta\s}}
{{h(\tau + \Delta\tau, \s) - h(\tau,\s)}\over{\Delta\tau}}]$$
$$=  f(\tau,\s) [
g(\tau + \Delta\tau, \s) 
h(\tau, \s + \Delta\s) - h(\tau + \Delta\tau, \s) g(\tau, \s + \Delta\s) ]$$
$$+ f(\tau, \s)g(\tau, s) [ h(\tau + \Delta\tau, \s) -h(\tau, \s + \Delta\s)]
- f(\tau, s) h(\tau,s)   [ g(\tau + \Delta\tau, \s) -g(\tau, \s + \Delta\s)].$$

If the four sides of this square are interpreted as a necklace for a closed string state whose beads are the four corners, the terms in $B$ can be expressed as 
\eqn\squarebeads{\langle f(\s_1) g(\s_2) h(\s_3) - f(\s_1) h(\s_2) g(\s_3)\rangle }
$$+ \langle f(\s_1) g(\s_1) (h(\s_2) - h(\s_3))\rangle
- \langle f(\s_1) h(\s_1) (g(\s_2) - g(\s_3))\rangle$$
$$=
\langle f(\s_1) g(\s_2) h(\s_3) - f(\s_1) h(\s_2) g(\s_3)\rangle$$
where the expression is assumed to depend only on the cyclic order of the beads ($\s_1\leq\s_2\leq\s_3\leq\s_4$) since the theory is topological when $\l_{tHooft}=0$. In other words, the factor of  $\p_\tau g \p_\s h - \p_\s g \p_\tau h$ in $B$ has turned into the commutator $[g,h]$ in the cubic vertex. This is reminiscent of the Poisson bracket \deWit\ which relates the supermembrane action and M(atrix) theory.

By expanding to lowest order in the worldsheet variables, evidence will now be presented that $B$ of \Bdef\ indeed generates the cubic term in the super-Yang-Mills action of \sym. Note that both $\int d^2 z ~B$ and the cubic super-Yang-Mills vertex are in the BRST cohomology and are $PSU(2,2|4)$ invariant, so showing equivalence at the lowest non-trivial order is strong evidence for equivalence to all orders in the worldsheet
variables. 

\subsec{ Expansion of B}

To compare with the cubic super-Yang-Mills vertex of \sym, it will be useful to expand $B$ to lowest order in the worldsheet variables. Although it might seem surprising that this expansion makes sense at zero radius, note that the parameter in front of the topological action is not the AdS radius and can be taken as large as desired.
In this limit, the first term in $B$ is
\eqn\nontriva{ {1\over 4}\eta_{\a\b} J_{1}^\a \wedge J_{3}^\b \sim {1\over 4} \eta_{\a\b} d\t_L^\a \wedge d\t_R^\b +{1\over 4} \g^a_{\a\b} x_a  (d\t_L^\a \wedge d\t_L^\b - d\t_R^\a \wedge d\t_R^\b) + ...}
$$\to {1\over 4}
 \g^a_{\a\b} x_a ( \{\t_L^\a ,\t_L^\b\} - \{\t_R^\a, \t_R^\b\})+ ...$$
where $...$ denotes terms higher-order in the worldsheet fields, total derivative terms are ignored, and the anti-commutators come from the discussion of the previous subsection. Similarly, the second term in $B$ is
\eqn\nontrivb{{{(\l_L \eta\g_{ab}\l_R) + (\l_L \g_{ab}\eta\l_R)}\over{8(\l_L\l_R)}} J_{2}^a \wedge J_{2}^b\sim 
{{(\l_L\eta\g_{ab}\l_R)+(\l_L \g_{ab}\eta\l_R)}\over{8(\l_L\l_R)}} [x^a , x^b] + ...,}
and the third term in $B$ is
\eqn\nontrivc{{{(\l_L \g^\a J_{1})\wedge (\l_L \eta \g_a \eta J_{1})+(\l_R\g^a J_3)\wedge (\l_R\eta\g_a\eta J_3)}\over{8(\l_L\l_R)}} \sim}
$$
{{(\l_L \g^\a)_\a (\l_L \eta \g_a \eta)_\b}\over{8(\l_L\l_R)}} \{\t_L^\a, \t_L^\b\} 
+ {{(\l_R \g^\a)_\a (\l_R \eta \g_a \eta)_\b}\over{8(\l_L\l_R)}} \{\t_R^\a, \t_R^\b\} +  ... $$
Note that these terms are invariant under constant shifts of $\t$ as expected because of supersymmetry.

So to lowest order in the worldsheet variables, the closed vertex operator is described by the sum of \nontriva, \nontrivb\ and \nontrivc, and evidence will now
be presented that these terms generate the cubic super-Yang-Mills vertex of \sym. The most direct method for showing this would be to compare these terms with the vertex of \varb\ where $E_1$, $E_2$ and $E_3$ are chosen to correspond to the three super-Yang-Mills fields in the vertex. But the explicit form of $E$ has not yet been worked out for the gluinos $\psi^\alpha$ or for the auxiliary fields $D_{ab}$. 

A more indirect method is to compute the open-closed amplitude defined in Figure 2 where the three open 
strings describe the super-Yang-Mills states in the cubic vertex and the closed vertex operator is $B$. Using the OPE's from the topological action of these open string vertex operators with the terms in the closed vertex operator, it should be possible to compute explicitly this disk amplitude. 
In the pure spinor formalism, the integrated vertex operator for the gluon $A_a$ and gluino $\psi^\a$ is 
\eqn\vgl{V_{open} =\int dz [ A_a (\p x^a + ...)  +{1\over 4}  F_{ab} ( ( w \g^{ab}\l) + ...) + \psi^\a (p_\a + ...) ]}
 where $w_\a$ is the conjugate momentum for $\l^\a$ and $p_\a$ is the conjugate momentum for $\t^\a$. So using the naive free field OPE's from flat space of this open
 string vertex operator with the closed vertex operator of $B$, one sees that the first term \nontriva\ in $B$ can generate
the cubic vertex $\g^a_{\a\b} A_a \{\psi^\a, \psi^\b\}$ and the second term \nontrivb\ in $B$ can generate the cubic vertex $F^{ab} [A_a, A_b]$. The third term \nontrivc\ in $B$ does not seem to contribute at this order to the cubic super-Yang-Mills vertex, but is needed for BRST invariance of the closed vertex operator. So evidence has been presented here using naive free field OPE's that the cubic super-Yang-Mills vertex is indeed generated by $B$, and it should be possible to confirm this in the near future by explicit computations using the topological action.

\newsec{Summary}

In this paper, the pure spinor worldsheet action for the $AdS_5\times S^5$ superstring at small radius was expanded around the topological action of \actionexp. At zero radius, it was argued by analyzing the structure of the pure spinor vertex operators that the string worldsheet splits into regions near the AdS boundary which describe the Feynman propagators connecting the free super-Yang-Mills states, and regions near the AdS horizon which describe the holes of the Feynman diagrams. And at small radius, evidence was presented that the deformation of the topological action described by an antisymmetric $B$ term reproduces the cubic super-Yang-Mills vertex of the Feynman diagrams, thereby proving the Maldacena conjecture at small `t Hooft coupling.

To convert these sketchy arguments into a rigorous proof of the conjecture, one would need to further study this topological action and develop methods for performing explicit computations. Although it appears to be consistent to allow inverse powers of the ghost combination $(\l_L \l_R)$ in order to define this topological action and compute topological amplitudes, it would be important to better understand the BRST cohomology associated to this enlarged Hilbert space and to show how to perform computations in the presence of these inverse powers of bosonic ghosts. It would also be interesting to relate these topological string computations with the usual string computations involving integrated vertex operators. A useful clue in determining the relation between the topological and usual string amplitude computations may come from the fact that $(\l_L\l_R)$ and $\int d^2 z B$ are respectively the unintegrated vertex operator and integrated vertex operator for the radius modulus.

\vskip15pt
{\bf Acknowledgements:}
I would
like to thank Andrei Mikhailov, Lubos Motl, Leonardo Rastelli, Warren Siegel, Cumrun Vafa, Herman Verlinde, Edward Witten and especially Juan Maldacena and Pedro Vieira
for useful discussions, and
CNPq grant 300256/94-9
and FAPESP grants 2016/01343-7 and 2014/18634-9 for partial financial support.

\newsec {Appendix: Review of Topological $AdS_5\times S^5$ Action}

In this appendix, the construction of \BerkovitsGA\ of the topological $AdS_5\times S^5$ action will be reviewed and a minor error will be corrected concerning a term in $B$.  Comments will then be made on the relation to a recent paper of Mikhailov \MikhailovMDO\ on a similar topological action.

For the $AdS_5\times S^5$ background, the worldsheet action is
\eqn\ssads{S = \int d^2 z [\half \eta_{ab} J_2^a \bar J_2^b - \eta_{\a\b} ({3\over 4} J_3^\a \bar J_1^\b +{1\over 4} \bar J_3^\a J_1^\b)}
$$
-w_{L\a}\bar\nabla \l_L + w_{R\a} \nabla \l_R - {1\over 8}\eta_{[ab][cd]} (w_L \g^{ab}\l_L)(w_R\g^{cd}\l_R) +
\eta^{\a\b} w^*_{L\a} w^*_{R\b}]$$
where $\eta_{[ab][cd]} = + \eta_{a[c}\eta_{d]b}$ if $(a,b,c,d)$ are in $AdS_5$, $\eta_{[ab][cd]} = - \eta_{a[c}\eta_{d]b}$ if $(a,b,c,d)$ are in $S^5$, and
$w^*_{L\a}$ and $w^*_{R\a}$ are fermionic antifields which are constrained to satisfy 
\eqn\antifie{w_{L\a}^* \g_a^{\a\b} \eta_{\b\g} \l_R^\g =0, \quad w_{R\a}^* \g_a^{\a\b} \eta_{\b\g} \l_L^\g =0}
and are introduced so that the BRST transformations are nilpotent offshell.

The nilpotent BRST transformations which leave \ssads\ invariant are 
\eqn\brstads{ Q J_2^a = (\l_L \g^a J_1) + (\l_R \g^a J_3), \quad Q J_1^\a = \nabla \l_L^\a -   (\l_R\g_a \eta)^\a J_2^a, \quad
Q J_3^\a = \nabla \l_R^\a + (\l_L \g_a \eta)^\a J_2^a, }
$$Q w_{L\a} = \eta_{\a\b} J_3^\b + w_{L\a}^*, \quad Q w_{R\a} = \eta_{\a\b} \bar J_1^\b + w_{R\a}^*, $$
$$
Q w^*_{L\a} = - \eta_{\a\b} {{\p L}\over {\p w_{R\b}}}, \quad Q w^*_{R\a} =  \eta_{\a\b} {{\p L}\over {\p w_{L\b}}}.$$

As discussed in \BerkovitsGA, one can define a topological $AdS_5\times S^5$ action as
\eqn\stopads{S_{top} = \int d^2 z Q(\Psi)}
$$=\int d^2 z [ {{(\l_L\eta \g_a \g_{b} \l_R) + (\l_L \g_a \g_b \eta\l_R)}\over{4(\l_L\l_R)}} J_2^{a} \bar J_2^{b} 
-  J_3 \eta \bar J_1$$
$$  - w_{L\a} \bar \nabla \l_L^\a + w_{R\a}\nabla \l_R^\a 
- {1\over 8}\eta_{[ab][cd]} (w_L \g^{ab}\l_L)(w_R\g^{cd}\l_R) +
\eta^{\a\b} w^*_{L\a} w^*_{R\b}$$
$$+{{(\l_L \eta \g^a \eta \bar J_1)(\l_L \g_a  J_1)}\over {4(\l_L\l_R)}}
- {{(\l_R \eta \g^a \eta J_3)(\l_R \g_a \bar J_3)}\over {4(\l_L\l_R)}}]$$
where\foot{The last line of \stopads\ and $B$ were mistakenly omitted in \BerkovitsGA.}
\eqn\psiads{\Psi = {1\over{4(\l_L\l_R)}} (\l_R \eta \g_a)^\a  (\half\bar J_2^a \eta J_3 
+ w_{L\a} (\l_L \g^a \bar J_1) ) -\half w_{L\a} \bar J_1^\a }
$$-  {1\over{4(\l_L\l_R)}} (\l_L \eta \g_a)^\a  (\half J_2^a \eta \bar J_1 - w_{R\a} (\l_R \g^a J_3) ) +\half w_{R\a} J_3^\a $$
$$+ \half \eta^{\a\b} (w_{L\a} w^*_{R\b} - w^*_{L\a} w_{R\b}).$$

Comparing \ssads\ with \stopads, one finds that the pure spinor worldsheet action in an $AdS_5\times S^5$ background can be expressed as $S = \int d^2 z [Q(\Psi) + B]$ where 
\eqn\bads{B ={1\over 4}(J_3 \eta \bar J_1 - \bar J_3 \eta J_1) - {{(\l_L\eta\g_{ab} \l_R) + (\l_L \g_{ab} \eta\l_R)}\over{4(\l_L\l_R)}} J_2^{a} \bar J_2^{b} }
$$- {{(\l_L \eta \g^a\eta \bar J_1)(\l_L \g_a J_1)}\over {4(\l_L\l_R)}}
+ {{(\l_R \eta \g^a \eta J_3)(\l_R \g_a \bar J_3)}\over {4(\l_L\l_R)}}$$
is antisymmetric under the exchange of $z$ and $\bar z$. Note that in the flat space limit,
$B$ of \bads\ reduces up to total derivatives to the BRST-closed expression
\eqn\bflat{B^{flat} =L_{WZ} - {{(\l_L\eta \g_{ab} \l_R) + (\l_L\g_{ab} \eta\l_R)}\over{4(\l_L\l_R)}}\Pi^{a} \bar \Pi^{b} }
$$+ {{(\l_L \eta \g^a d_R)(\l_L \g_a \p\t_L)}\over {4(\l_L\l_R)}}
+ {{(\l_R \eta \g^a d_L)(\l_R \g_a \bar\p\t_R)}\over {4(\l_L\l_R)}}$$   
where $\Pi^a$ and $d_\a$ are the usual supersymmetric bosonic and fermionic momenta in flat space and $L_{WZ}$ is the antisymmetric Wess-Zumino term in the Green-Schwarz flat space action.

In a recent paper \MikhailovMDO, Mikhailov considered a slightly different construction of the $AdS_5\times S^5$ topological action $S= \int d^2 z [Q(\tilde\Psi) + \tilde B]$ where 
\eqn\mikpsi{\tilde\Psi = \Psi +   {1\over{8(\l_L\l_R)}} (\l_R \eta \g_a \eta)_\a   J_2^a \wedge  J_3^\a +{1\over{8(\l_L\l_R)}} (\l_L \eta \g_a\eta)_\a  J_2^a \wedge J_1^\a,}
\eqn\mikB{\tilde B = B -{{(\l_L \g^a J_1) \wedge (\l_R \eta\g_a \eta J_3)}\over {4(\l_L\l_R)}} 
+ {{(\l_L\eta\g_{ab} \l_R) + (\l_L \g_{ab} \eta\l_R)}\over{8(\l_L\l_R)}} J_2^a\wedge  J_2^{b}}
$$- {{(\l_L \eta \g^a \eta J_1) \wedge(\l_L \g_a J_1)}\over {8(\l_L\l_R)}}
- {{(\l_R \eta \g^a \eta J_3)\wedge (\l_R \g_a  J_3)}\over {8(\l_L\l_R)}}$$
\eqn\btilde{= {1\over 4}\eta_{\a\b} J_3^\a \wedge J_1^\b  - {{(\l_L \g^a J_1) \wedge (\l_R \eta\g_a \eta J_3)}\over {4(\l_L\l_R)}}}
and $\Psi$ and $B$ are defined in \psiads\ and \bads. 
Although the expression of $\tilde B$ in 
\btilde\ is simpler than the expression for $B$ in \bads, the topological construction of \MikhailovMDO\ has the disadvantage that it cannot be obtained by deforming the flat space worldsheet action. In other words, $\tilde B$ of 
\btilde\ has no flat space limit analogous to \bflat\ which is BRST-closed and contains the usual Green-Schwarz Wess-Zumino term $L_{WZ}$.

\listrefs

\ifig\figtwelve{  3-point amplitude on sphere where orange circles are closed string vertex operators, blue strips are thickened propagators near the AdS boundary, white regions are D3-brane holes near the AdS horizon, black dots are picture-raising operators, and red dots are “beads” E on the closed strings   }
{\epsfxsize 3 in\epsfbox{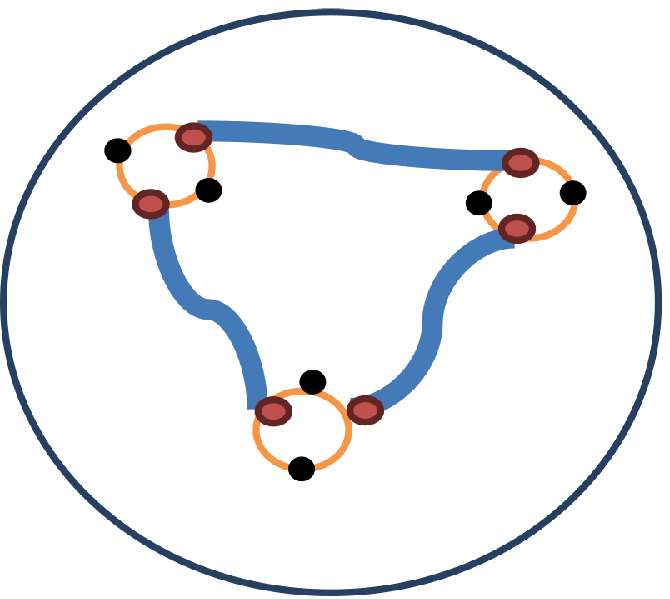}}

\ifig\figthirteen{ Open-closed amplitude on disk where purple lines are open string vertex operators   }
{\epsfxsize 3 in\epsfbox{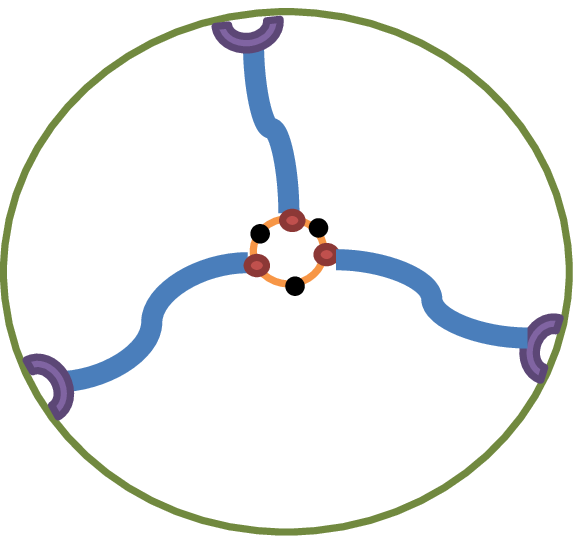}}

\end